# Procesamiento Digital De Imágenes Aplicado A La Segmentación De Objetos Por intensidad y movimiento.

Benjamín Andrés Huérfano Zapata[1] Universidad de Cundinamarca

*Resumen*— El desarrollo tecnológico actual nos permite llevar a cabo tareas que hace algún tiempo eran impensables por no decir que imposibles, el procesamiento digital de imágenes ha sido una de las mayores constantes de desarrollo en la actualidad, teniendo en cuenta que su implementación data de poco tiempo atrás, OpenCV [1] es una herramienta enfocada en la visión artificial, en este caso implementada en una plataforma de programación orientada a objetos basado en lenguaje en Java[1] ofrecida por el software de desarrollo NetBeans[2], Teniendo como base lo anterior se planteó e implemento una plataforma física a modo de ambiente cerrado la cual a través del desarrollo de un algoritmo permitió detección y segmentación de objetos por medio del modelo de color RGB; en trabajos futuros dicho algoritmo brindara la base de información para la plataforma robótica autónoma; este avance abre un amplio espectro para el desarrollo de aplicaciones y herramienta en el campo de la visión artificial.

*Palabras clave*— Detección de objetos, RGB, Visión artificial, OpenCV, C#, EmguCV.

## I. INTRODUCCIÓN

El avance tecnológico actual ha venido presentando una constante evolución en cuanto al desarrollo de software se refiere como lo señala Gonzales [2] esto ha traído consigo numerosas ventajas, entre las cuales cabe destacar el desarrollo de tecnologías que son por mucho más eficientes, rápidas y baratas [3]; como resultado de lo anterior se hace plausible e interesante abordar ciertas temáticas que hace algún tiempo pudieron haber sido despreciadas o desechadas por la exigencia de gran cantidad recursos computacionales. Gracias a que las nuevas tecnologías han conseguido mejorar en gran medida las velocidades de procesamiento de la información y así un vasto camino para estudio en el desarrollo de algoritmos mucho más eficaces.

El actual documento expone el desarrollo de un algoritmo basado en programación orientado a objetos basado en el lenguaje Java, a través de la plataforma de programación brindada por el software NetBeans, que en conjunto y haciendo uso de la librería OpenCV de visión artificial desarrollada por Intel, la cual entre otra ventajas cuenta con licencia libre desde su aparición con la primera versión alfa en el mes de enero de 1999, ha sido implementada en infinidad de aplicaciones: desde sistemas de seguridad con detección de movimiento, hasta aplicaciones de control de procesos en los que se requiere reconocimiento de objetos. Esto se debe a que su publicación se da bajo licencia BSD, que permite que sea usada libremente para propósitos comerciales y de investigación con las condiciones en ella expresadas.

## II. DESCRIPCIÓN Y CARACTERIZACIÓN DEL DISEÑO DEL ALGORITMO:

En el actual documento plantea como problemática esencial la detección y segmentación de varios objetos de colores previamente determinados, en un entorno compuesto distintos objetos para esto se usó una descomposición de la imagen pixel por pixel y un posterior filtrado de histogramas por ecualización. El algoritmo debía de estar en la capacidad de identificar la distancia y el centro del objeto (de un color en específico) con respecto a los demás objetos y así mismo mostrar lo resultados obtenidos de la imagen, además de segmentar e identificar cada objeto como independiente sin importar si contaban con el mismo color.

Para lo cual se identificaron y definieron aspectos generales a abordar para esta manera poder implementar un algoritmo mucho más robusto y completo.

### A. Caracterización del ambiente de trabajo:

Se procede a identificar las características básicas esenciales del ambiente de estudio, el cual lo enmarcaremos con los rasgos primordiales de una área real de trabajo de tipo industrial automatizado [4], donde dicha área se caracteriza por contar con características de limpieza, no intersección de diferentes zonas de trabajo, ambiente propicio para el control implementado, entre otras cuantas.

Se definieron como características indispensables para un ambiente propicio un fondo monocromático que permitiera identificar los objetos, definido con color blanco el fondo implementado cuneta con un área de $1m^2$ con longitudes laterales equitativas de $100cm$.

Ubicación de la cámara frontal a fin de lograr un mapeo general del fondo usado.

Ambiente con buenas condiciones de iluminación la cual se implementó iluminación de tipo frontal la cual ayuda permite la iluminación uniforme y es usada comúnmente en sistemas de detección de objetos [5].

### B. Modo de adquisición de información:

La adquisición de información se realizara a través de una cámara HD (Logitech HD Pro Webcam C920) por medio de protocolo USB 2.0, ofreciendo de este modo gran calidad de imagen y eficacia en la trasmisión de información [6].

La cual se encuentra situada a $154cm$ con el fin de permitir un pixel para la resolución actual represente un área $2.25mm^2$, la cual fue definida por el área de cobertura proporcionada por la cámara que conto con $96cm$ de largo y $72cm$ de ancho para un total $0.69m^2$.

---

[1] "Java es un lenguaje de programación y una plataforma informática comercializada desde en 1995 por Sun Microsystems" [10]. La versión de Java usada actualmente es la orientada a objetos.

[2] NetBeans es un proyecto de código abierto dedicado a proporcionar productos de desarrollo de software solidos como NetBeans IDE y NetBeans Platform [11].

### C. Selección de escala del color:

El desarrollo expuesto por el presente se basó en la creación un algoritmo, el cual permitió identificar los colores primarios en este caso en la escala de RGB[3] (Rojo, Verde Y Azul) puesto que el modelo de codificación usado por la cámara seleccionado es el anteriormente mencionado, y con el fin de evitar pérdidas de recursos computacionales y eficiencia del algoritmo en la trasformación a otro tipo de escala.

### D. Tiempo de muestreo:

El tiempo de muestreo como meta definido para el actual es definido el fin de contar con una continuidad mínima de fotogramas, cercana a la imperceptible por el ojo humano, por otra parte dicha cantidad de muestras por unidad de tiempo está limitada por la cantidad posible de obtener por la cámara seleccionada siendo un máximo de $30 fps$.

Finalmente teniendo en cuenta una cantidad de tiempo prudente de procesamiento de la mitad del tiempo de captura dando un margen meta de mínimo $15 fps$.

### E. La interfaz gráfica:

La interfaz gráfica debía de contar con característica de modularidad con el fin de permitir identificar con facilidad los resultados obtenidos y poder mantener de forma ordenada cada uno de sus componentes.

Debía contar con la posibilidad de realizar ajustes sobre la ejecución para diferentes variables de importancia dentro del algoritmo, igualmente contar con la posibilidad de observar los resultados obtenidos de forma tanto visual como matemática.

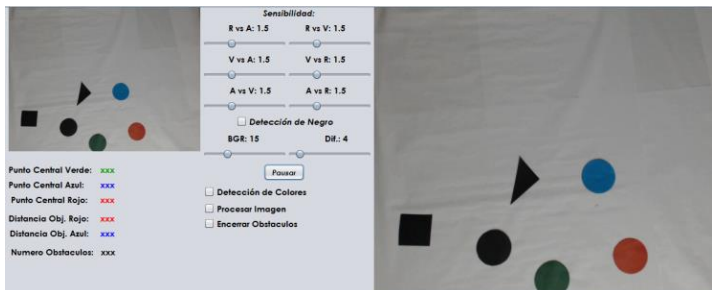

**Figura 1 –** Interfaz grafica implementada
**Fuente:** Autor

### F. Caracterización de los objetos:

Los objetos inmersos dentro del ambiente contaban con características diferenciables a primera vista con el fin de poder leer los resultados con mayor facilidad para el actual documento. Por tanto definimos los colores Verde, Rojo, Azul y negro, dichos colores serán los que se segmentaran de nuestro ambiente de trabajo.

Definimos un tamaño mínimo de $25 cm^2$ para cualquier objeto inmerso dentro del ambiente, con el fin de evitar que la aparición de ruido inmerso en la imagen llegase a afectar las mediciones obtenidas.

## III. MODO DE ALMACENAMIENTO DE LA IMAGEN

Para llegar a al desarrollo del actual algoritmo se debe esclarecer previamente sobre la forma en la que se almacena la información de la imagen. Aunque una imagen es la reproducción de la figura de un objeto por la combinación de los rayos de luz que inciden en el mismo, dentro de un ordenador una imagen no es más que una gran secuencia de datos. Como es conocido, el tamaño de las imágenes se mide en píxeles, que es precisamente la superficie homogénea más pequeña de las que componen una imagen [7], que se define por su brillo y color lo cuales esta definidos por el valor que toma el pixel.

Las imágenes en blanco y negro (binarias) son las más básicas, pudiendo tomar sus píxeles valores entre 0 (completamente negro) y 1 (completamente blanco). Sin embargo las imágenes de color contienen tres canales de colores distintos (Para nuestro caso RGB por sus siglas en inglés (R : rojo, G : verde, B : Azul) . Cada canal puede tomar un valor entre 0 y 255 correspondiente a un byte de información, por ejemplo un $R = 255, G = 0,$ y $B = 0$ será un píxel completamente rojo. Si los tres valores son 255 el píxel será blanco, y si los tres valen 0 el píxel será negro, una vez definido esto se identifica como a través de la mezcla de estos tres colores base se obtiene $16'581.375$ posibilidades por pixel, permitiendo así tener una gran cantidad de matices.

La imagen ademas se compone de caracteristicas principales tales como altura, ancho, numero de canales. Por tanto para recorrer los valores de una imagen se puederelizar de izquierda a derecha o de arriba a abajo. Teniendo en cuenta además que los canales están ordenados en BGR (azul, verde y rojo). Está es la única forma de recorrer imágenes, y se implementa con bucles anidados aplicados a una matriz o array dependiendo el modo de almacenamiento o de obtencion de informacion usado.

## IV. IDENTIFICACIÓN POR COLOR

Para detectar un objeto de determinado color inicialmente se necesita identificar todos los píxeles que lo componen. Para ello se realiza una búsqueda en todos los datos de la imagen, calculando si son o no del color deseado. Para la distinción del color que se está evaluando en ese instante; se realiza, en primer lugar la selección de datos característicos que contienen el pixel de color y finalmente la cercanía con el color deseado, con el fin de clasificarlo o no dentro del tipo de color buscado. Por tanto su dato más característico será que el canal con mayor valor. Es de vital importancia tener en cuenta que los valores de los demás canales no deberían ser muy altos, por lo menos no lo suficiente como para acercarse al valor del canal testeado, ya que estos valores podrían corresponder a colores como el morado, el naranja, el amarillo entre otros, lo cual crearía lectura erróneas sobre la captura de la imagen y arrojarían resultados indeseados, con este fin de delimito el algoritmo a partir de los canales del pixel evaluado, mirando que el valor del pixel estaba contenido entre el espectro de color que deseamos identificar.

---

[3] RGB es un modelo de color basado en la síntesis aditiva, con el que es posible representar un color mediante la mezcla por adición de los tres colores de luz primarios [12].

Con un color que no pertenece de los tres básicos de RGB (Ej. negro), se vuelve un poco más complejo este paso, pues se debe definir unas restricciones más estrictas y un rango amplio para el cual se puede clasificar el color, de igual manera al tener dichos cambios permitirá poder detectar los colores sin que los afecte la iluminación con tanta facilidad puesto las tonalidades y degrades presentes en la superficie del objetos se encontrarían inmersas en los mismos. La detección de colores resultantes de mezclas se debe identificar y clasificar en una matriz o de manera homologa una base de datos, permitiendo así acceder al valor y compararlo con el que se observa.

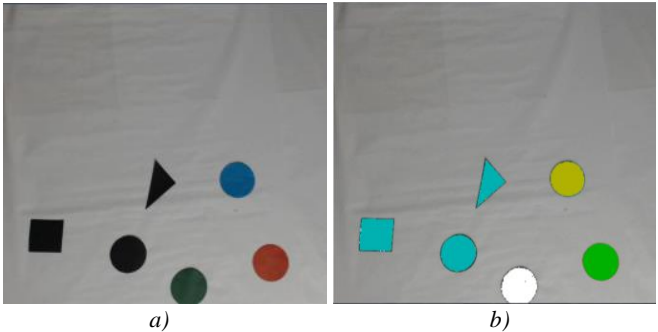

**Figura 2** – *a)* Imagen capturada y *b)* Imagen procesada por color.
**Fuente:** Autor.

La dificultad de este método es que solo detectaría colores de tipo solido frente al objeto testeado, sabiendo que la forma irregular de los objetos que puede presentar en la imagen nos puede efectos de cambio de la intensidad del color del pixel, este caso lo anterior se soluciona dejando un rango neutro adicional entre los colores mezcla y los RGB primarios, y a la par utilizar un extrapolación de la imagen para detección de bordes con el uso de primeras y segundas derivadas entre los valores de lo pixeles adyacentes [8], lo cual nos permitiría los cambios brucos de entorno y así mismo determinar los bordes del objeto deseado.

## V. POSICIONAMIENTO

El posicionamiento permitirá adquirir información vital del ambiente en el cual se está desarrollando el estudio, en este caso hace referencia al posicionamiento de los objetos identificados, permitiendo identificar características como tamaño, forma, distancia y centro del objeto estudiado.

Para ubicación de los objetos y de los pixeles de color deseado, se guardaron todas las coordenadas de cada pixel identificado y que se encontraban a su vez clasificado en los colores deseados (Rojo, Verde, Azul y Negro) gracias a la suma de sus posiciones, se calculó sería el punto medio de nuestro objeto a través de la siguiente formula:

$$\frac{\sum_{i \in OD}^{n} i}{\sum_{i \in OD}^{n} 1} \quad y \quad \frac{\sum_{j \in OD}^{m} j}{\sum_{i \in OD}^{m} 1}$$

Donde $n$ y $m$ representa el largo y ancho de la imagen en pixeles respectivamente, $OD$ el conjunto de pixeles de un objeto de un color deseado. Con estos se consigue la coordenada en $(x,y)$ en pixeles del centro del objeto evaluado, lo cual posteriormente nos servirá de base para el cálculo de distancia entre estos objetos y de igual forma para la segmentación de objetos. Estás coordenadas del punto medio serán bastante precisas si contamos con una gran cantidad de píxeles del objeto, y nos será de gran utilidad para ubicar dicho objeto con respecto a nuestra plataforma robótica.

Supondremos a contamos con un robot un punto de referencia de un color este caso se asigna el color verde. Calculamos el punto medio de dicho objeto de referencia, Una vez tenemos las coordenadas sabemos si el objeto se encuentra a nuestra izquierda o derecha, y arriba o abajo de nuestra plataforma robótica un punto de llegada en la zona de color rojo o azul. Procedemos a hallar la distancia del robot con respecto a las zonas de llegada (en píxeles) mediante la fórmula de la hipotenusa:

$$Distancia = \sqrt{(x_2 - x_1)^2 + (y_2 - y_1)^2}$$

Donde $x_i$ representa la coordenada $x$ y $y_i$ la coordenada en $y$ dados pixeles del objeto $i$.

| Punto Central Verde: | 296, 453 |
|---|---|
| Punto Central Azul: | 373, 283 |
| Punto Central Rojo: | 427, 415 |
| Distancia Obj. Rojo: | 136px 204 mm |
| Distancia Obj. Azul: | 142px 213 mm |

**Figura 3** – Posicionamiento de objetos en la *figura 3 b)*.
**Fuente:** Autor.

Ahora contamos con la posición y distancia entre objetos, pero como lo indica la *figura 4* aun no conocemos el área de la imagen que ocupa cada objeto se conoce que ($\sum_{i \in OD}^{n} 1 * 2.25 mm^2$) nos indica el área real aproximada ocupada por el objeto actual. Pero debido a la naturaleza de los objetos que es irregular se hace necesario enmarcar el objeto en un área cuadrada un poco más amplia para contar con un margen de distancia entre un obstáculo y lo definido como plataforma robótica para el documento, para efectos del estudio definiremos el color negro como obstáculos, los cuales serán enmarcados de la manera anteriormente propuesta, por tanto para lograrlo se realiza un sondeo de la coordenada mínima y máxima del obstáculo evaluado del siguiente modo:

$$\left(\min_{i \in OD} i, \min_{j \in OD} j\right) \quad y \quad \left(\max_{i \in OD} i, \max_{j \in OD} j\right)$$

Proporcionándonos de este modo un área cuadrada cual enmarca el obstáculo que estudiado.

Finalmente usando a forma de un plano cartesiano la imagen adquirida se ubican en dos dimensiones los objetos lo cual brinda una perspectiva panorámica de ambiente en el que se encuentra el robot (color verde).

## VI. SEGMENTACIÓN

La segmentación se centra en cómo aislar los objetos o partes de objetos del resto de la imagen. Las razones para hacer esto son obvias. Por ejemplo, una cámara suele mirar hacia el mismo fondo, que no es de ningún interés. Lo que interesa de esa imagen es detectar cuándo las personas o vehículos entran en la escena, o cuándo algo se deja en la escena que no estaba allí antes [9].

Por tanto se desea aislar esos acontecimientos e ignorar interminables horas en que nada está cambiando. Este es un ejemplo de los muchos usos que tiene la segmentación de objetos en el procesamiento de imágenes.

La segmentación de los objetos consiste en como su nombre lo indica, en una división o separación de objetos con el fin de indicar que son distintos y que se encuentran en otra parte del plano al que se está observando. Además permite tener una mayor perspectiva del número de objetos al que se observa y asimismo generar un posicionamiento de cada uno de ellos con respecto a la plataforma robótica. En este caso supusimos es el objeto de color verde.

La segmentación de los objetos que son objeto de estudio se realizó por medio de las diferencia entre distancias de distintas aglomeraciones de pixeles perteneciente a obstáculos, para lo cual haremos uso de un rango definido de $10\ px$, donde una diferencia entre el ultimo de pixel $\in OD$ de la imagen con respecto al siguiente en una misma fila o columna de la imagen, representara un nuevo posible objeto, si la nueva aglomeración de pixeles encontradas supera el área mencionada en $(II.F)$ se considera como un objeto nuevo y distinto al primer encontrado.

Logrando así detectar múltiples objetos del mismo color y que en conjunto a las características expresada en $(III.)$ permiten su total segmentación y ubicación. En este caso serán los objetos definidos como obstáculos anteriormente (color negro) que estarán en la misma imagen pero diferentes áreas de la imagen.

En este caso sabemos que hay distintos objetos que identificados, y que unos píxeles pertenecerán a un obstáculo, otros píxeles pertenecerán a otros. ¿Cómo saber de qué objeto se trata? Para ello haremos uso de una instancia del tipo de programación que estamos empleando es la creación de entidades conocidas como objetos, los cuales permitirán guardar la información de cualquier obstáculo presente en el área de estudio. Obteniendo finalmente lo representado por la *figura 5*.

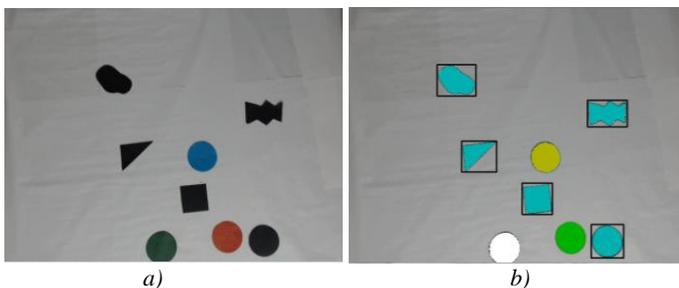

**Figura 4 –** *a)* Imagen capturada y *b)* Imagen segmentada.
**Fuente:** Autor.

Donde se puede observar como las características anteriormente mencionadas se reflejan en la *figura 5 b)*, donde los obstáculos sin importar su forma son enmarcados e identificados como distintos. Sin ser afectados por la cercanía con otro objeto de distinto color, los cuales siguen siendo posicionados y ubicados plenamente.

## VII. IMPLEMENTACIÓN DEL ALGORITMO

Tomando lo anteriormente expuesto se procedió a generar el código del algoritmo el cual conto con una estructura como la expuesta por la *figura 6*:

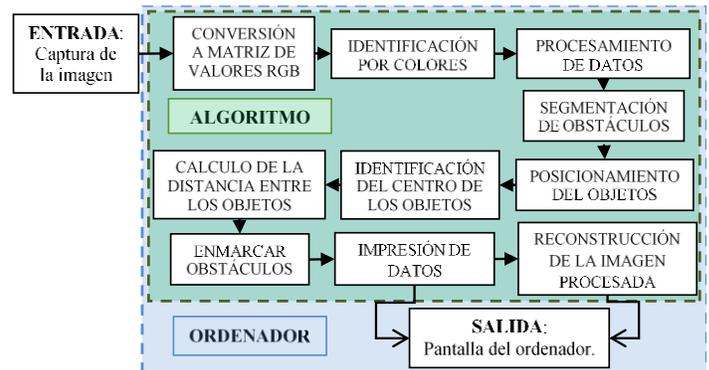

**Figura 5 –** Diagrama de bloques del algoritmo.
**Fuente:** Autor.

En detalle el algoritmo realiza se presenta como:

- Inicialización el programa y definición de variables globales y especiales.
- Para la carga de la imagen obtenida a través de la cámara por medio de protocolo USB, se tienen en cuenta los posibles errores de carga y dificultades del ordenador y se generan las excepciones necesarias con el fin de evitar fallos en el programa y en el computador.
- Se obtienen las características de la imagen datos con los cuales podremos recorrer la imagen.
- Convertimos le mapa de bits en una matriz la cual se recorre para acceder a los valores de los píxeles teniendo cuenta sus dimensiones y características obtenidas del punto inmediatamente anterior.
- Se evalúa cada valor y se clasifica el pixel según las restricciones y rangos anteriormente dispuestos.
- Se modifican los pixeles dentro de la matriz parar indicar el cambio detección del color.
- Se evalúa la presencia de posibles objetos del mismo color.
- Se genera segmentación de obstáculos de modo indicado en el apartado $(VI.)$.
- Por medio de las ecuaciones mencionadas en el apartado de posicionamiento se realizan los respectivos cálculos de centro de cada objeto y medida de distancias.
- Se enmarcan los obstáculos segmentados identificados.
- Y finalmente se recrea la imagen y se muestran los resultados obtenidos por el algoritmo. De manera seguida se liberan recursos del programa Se regresa al típico inicial.

## VIII. ANÁLISIS Y RESULTADOS

Una vez desarrollado el algoritmo se procedió a realizar el testeo del mismo, el cual se realizó en un computador personal de gama media con las característica indicadas en el *cuadro 1*.

| Procesador: | Intel Pentium $2.2\ GHz$ |
|---|---|
| **Memoria RAM:** | $6\ GB$ |
| **Tarjeta de video:** | Intel(R) HD Graphics integrada |
| **Cámara:** | Logitech HD Pro Webcam $C920$ |
| **Sistema operativo:** | Windows 7 Ultimate – $64\ bits$ |

| NetBeans IDE: | 8.1 |
|---|---|
| Java(TM): | $1.8.0\_45 - b15$ |
| OpenCV | 3.0.0 |

**Cuadro 1** – Características de hardware y de software de ordenador usado.

Para realizar las pruebas se tomos una contante de 10 medidas las con aras de tener un buen margen acercamiento al valor real de fps y frecuencia de muestreo logrado por el algoritmo, dichas medidas se realizaron por medio de un código en Java el cual mostraba la hora de inicio de ejecución, la hora el final y el número de fotogramas alcanzados a procesar a como punto inicial tomaremos la los datos sin la aplicación del algoritmo únicamente la muestra de imágenes a través de OpenCV y Java como lo ilustra el *cuadro 2*.

| Tiempo de ejecución ($Seg$) | No. Imágenes procesadas | FPS ($Hz$) | Muestreo ($Seg$) |
|---|---|---|---|
| 1 | 21 | 21 | 0,0476 |
| 1 | 20 | 20 | 0,0500 |
| 2 | 44 | 22 | 0,0455 |
| 3 | 84 | 28 | 0,0357 |
| 2 | 47 | 23,5 | 0,0426 |
| 2 | 42 | 21 | 0,0476 |
| 3 | 66 | 22 | 0,0455 |
| 3 | 72 | 24 | 0,0417 |
| 4 | 78 | 19,5 | 0,0513 |
| 5 | 122 | 24,4 | 0,0410 |

**Cuadro 2** – Resultados sin aplicación del algoritmo.

Por tanto se identifica que la frecuencia promedio es de $22.54 fps$ lo cual nos deja un margen pequeño de trabajo de 7.54fps menos para lograr el margen esperado. Seguidamente se identificó el comportamiento de parte del algoritmo por tanto se procedió a realizar la prueba únicamente con la detección de color, con el fin de identificar la carga computacional que esta tarea conlleva para el actual algoritmo.

| Tiempo de ejecución ($Seg$) | No. Imágenes procesadas | FPS ($Hz$) | Muestreo ($Seg$) |
|---|---|---|---|
| 2 | 43 | 21,5 | 0,0465 |
| 2 | 44 | 22 | 0,0455 |
| 3 | 49 | 16,3 | 0,0612 |
| 3 | 48 | 16 | 0,0625 |
| 4 | 74 | 18,5 | 0,0541 |
| 4 | 71 | 17,7 | 0,0563 |
| 4 | 73 | 18,2 | 0,0548 |
| 3 | 68 | 22,7 | 0,0441 |
| 6 | 102 | 17 | 0,0588 |
| 3 | 59 | 19,7 | 0,0508 |

**Cuadro 3** – Resultados identificación por color únicamente.

Dando un promedio de $19,97 fps$ dando un costo de procesamiento de $3,57 fps$ lo cual indica que el costo de procesamiento por imagen aumento un 15,84%, dejando un margen con respecto a al margan esperado de $3,93 fps$. Ahora con la inclusión del método de segmentación expuesto en ($VI.$) se obtiene.

| Tiempo de ejecución ($Seg$) | No. Imágenes procesadas | FPS ($Hz$) | Muestreo ($Seg$) |
|---|---|---|---|
| 3 | 46 | 15,3 | 0,0652 |
| 2 | 41 | 20,5 | 0,0488 |
| 3 | 49 | 16,3 | 0,0612 |
| 3 | 53 | 17,6 | 0,0566 |
| 3 | 48 | 16 | 0,0625 |
| 2 | 42 | 21 | 0,0476 |
| 2 | 46 | 23 | 0,0435 |
| 5 | 82 | 16,4 | 0,0610 |
| 5 | 73 | 14,6 | 0,0685 |
| 2 | 40 | 20 | 0,0500 |

**Cuadro 4** – Resultados Identificación por color y segmentación de obstáculos.

Con un promedio de $18,08 fps$ presentando una pequeña reducción en comparación con lo avalado anteriormente con solo $0,89 fps$ esto debido a que los cálculos realizados dependen de valores anteriormente encontrados por al algoritmo y son pequeños en comparación además de que depende directamente del número de obstáculos.

| Tiempo de ejecución ($Seg$) | No. Imágenes procesadas | FPS ($Hz$) | Muestreo ($Seg$) |
|---|---|---|---|
| 5 | 76 | 14,6 | 0,0685 |
| 4 | 66 | 16,5 | 0,0606 |
| 3 | 51 | 17 | 0,0588 |
| 4 | 65 | 16,2 | 0,0615 |
| 4 | 70 | 18 | 0,0571 |
| 3 | 41 | 14 | 0,0732 |
| 4 | 64 | 16 | 0,0625 |
| 4 | 55 | 13,7 | 0,0727 |
| 3 | 60 | 20 | 0,0500 |
| 7 | 141 | 20 | 0,0496 |

**Cuadro 5** – Resultados algoritmo final.

Finalmente se realiza el análisis del algoritmo completo con la inclusión del posicionamiento y la identificación del área del objeto como lo expuesto por la *figura 5,* se obtiene un promedio de 16,54fps dejando un parte positivo en el rendimiento promedio esperado del algoritmo. Cabe mencionar que las distintas variaciones de las frecuencias de muestreo se deben a que al ser una seria distintas de imágenes, objetos y obstáculos, lo cual implica diferentes cargas computacionales.

## IX. CONCLUSIONES

- La implementación y diseño del algoritmo que nos permita procesamiento de imágenes usando librerías OpenCV aplicados a la detección de colores RGB, para la detección de objetos sobre una plataforma robótica muestra un gran eficacia gracias condiciones robustas de trabajo y grandes delimitaciones del mismo.

- La detección y segmentación de objetos por medio del procesamiento de imágenes es de gran utilidad para la obtención de datos del ambiente en el que se esta realizando la tarea de captura de imágenes, puesto nos muestra de forma muy directa la conexión entre los objetos y su centros.

- Aunque el ambiente recreado para el estudio no contaba con todas las condiciones necesarias de un ambiente industrial óptimo para el desarrollo de procesamiento de imagenes el algoritmo permitió resolver en su totalidad el problema planteado sin traer repercusiones a los tópicos planteados inicialmente (*II.*).

- La velocidad con la que arroja los resultados en el análisis de tiempo real nos permite tener un flujo de imágenes de más de $15 fps$ sin afectar el rendimiento del algoritmo, lo cual deja claro la calidad de ejecución y detección de objetos del mismo.

- El algoritmo es capaz de identificar, posicionar, medir y segmentar los elementos dentro de una ambiente cerrado con las características planteadas en el actual documento con gran eficacia con un 100% de los casos la detección, segmentación y ubicación de los elementos se llevó a cabo, en comparación con su precisión que fue exiguamente menor con un 96.71% con respecto a los datos cotejados de manera directa con una precisión 95.08% y 98.34% para medidas de punto medio de los objetos y distancia entre los mismos respectivamente.